\documentstyle[12pt,epsfig]{article}
\begin{document} \setlength{\baselineskip}{5mm}

\vbox{\vspace{20mm}}

\begin{center}

{\Large \bf Internal Space-time Symmetries of Massless Particles
and Neutrino Polarization as a Consequence of Gauge Invariance}\\[4mm]

{\large \bf Y. S. Kim} \\[4mm]

{\it Department of Physics, University of Maryland, College Park,
  Maryland 20742, U.S.A.}

\end{center}

\vspace{10mm}

\begin{abstract}
There are gauge-transformation operators applicable to massless
spin-1/2 particles within the little-group framework of internal
space-time symmetries of massive and massless particles.
It is shown that two of the $SL(2,c)$ spinors are invariant under
gauge transformations while the remaining two are not.  The Dirac
equation contains only the gauge-invariant spinors leading to
polarized neutrinos.  It is shown that the gauge-dependent $SL(2,c)$
spinor is the origin of the gauge dependence of electromagnetic
four-potentials.  It is noted also that, for spin-1/2 particles,
the symmetry group for massless particles is an
infinite-momentum/zero-mass limit of the symmetry group for
massive particles.

\end{abstract}

\vspace{30mm}

This note is based on one of the lectures delivered at the Advanced Study
Institute on Symmetries and Spin (Prague, Czech Republic, July 2001).

\newpage

\section{Introduction}\label{intro}
The internal space-time symmetries of massive and massless particles
are dictated by the little groups of the Poincar\'e group which are
isomorphic to the three-dimensional rotation group and the two-dimensional
Euclidean group respectively~\cite{wig39}.  The little group is the
maximal subgroup of the Lorentz group whose transformations leave the
four-momentum of the particle invariant.  Using the properties of these
groups we would like to address the following questions.

On massless particles, there are still questions for which answers are
not readily available.  Why do spins have to be parallel or anti-parallel
to the momentum?  While photons have a gauge degree of freedom with two
possible spin directions, why do massless neutrinos have only one spin
direction without gauge degrees of freedom?  The purpose of this note is
to address these questions within the framework of Wigner's little groups
of the Poincar\'e group~\cite{wig39}.

The group of Lorentz transformations is generated by three rotation
generators $J_{i}$ and three boost generators $K_{i}$.  They satisfy
the commutation relations:
\begin{equation}\label{eq1}
[J_{i}, J_{j}] = i\epsilon_{ijk} J_{k} , \qquad
[J_{i}, K_{j}] = i\epsilon_{ijk} K_{k} , \qquad
[K_{i}, K_{j}] = -i\epsilon_{ijk} J_{k} .
\end{equation}
In studying space-time symmetries dictated by Wigner's little group,
it is important to choose a particular value of the four-momentum,
For a massive point particle, there is a Lorentz frame in which the
particle is at rest.  In this frame, the little group is the
three-dimensional rotation group.  This is the fundamental symmetry
associated with the concept of spin.

For a massless particle, there is no Lorentz frame in which its momentum
is zero.  Thus we have to settle with a non-zero value of the momentum
along one convenient direction.  The three-parameter little group
in this case is isomorphic to the $E(2)$ group.  The rotational
degree of freedom corresponds to the helicity of the massless particle,
while the translational degrees of freedom are gauge degrees of
the massless particle~\cite{kiwi90,ino53}.

In this report, we discuss first the $O(3)$-like little group for a
massive particle.  We then study the $E(2)$-like little group for
massless particles.  The $O(3)$-like little group is applicable to
a particle at rest.  If the system is boosted along the $z$ axis, the
little group becomes a ``Lorentz-boosted'' rotation group, whose
generators still satisfy the commutation relations for the rotation
group.  However, in the infinite-momentum/zero-mass limit, the
commutation relation should become those for massless particles.
This process can be carried out for both spin-1 and spin-1/2 particles.
This ``group-contraction'' process in principle can be extended all
higher-spin particles.  In this report, we are particularly interested
in spin-1/2 particles.

In Sec. \ref{o3toe2}, we study the contraction of the $O(3)$-like
little group to the $E(2)$-like little group.  In Sec. \ref{massless},
we discuss in detail the $E(2)$-like symmetry of massless particles.
Secs. \ref{spinhalf} and \ref{gauge} are devoted to the question of
neutrino polarizations and gauge transformation.

\section{Massless Particle as a Limiting Case of Massive
Particle}\label{o3toe2}
The $O(3)$-like little group for a particle at rest is generated by
$J_{1}, J_{2}$, and $J_{3}$.  If the particle is boosted along the $z$
direction with the boost operator
$B(\eta) = \exp{\left(-i\eta K_{3}\right)}$,
the little group is generated by $J'_{i} = B(\eta)J_{i}B(-\eta)$.
Because $J_{3}$ commutes with $K_{3}$, $J_{3}$ remains invariant under
this boost. $J'_{1}$ and $J'_{2}$ take the form
\begin{equation}\label{eq3}
J'_{1} = (\cosh\eta)J_{1} + (\sinh\eta)K_{2} , \qquad
J'_{2} = (\cosh\eta)J_{2} - (\sinh\eta)K_{1} .
\end{equation}
The boost parameter becomes becomes infinite if the mass of the particle
becomes vanishingly small.  For large values of $\eta$, we can consider
$N_{1}$ and $N_{2}$ defined as $N_{1} = -(\cosh\eta)^{-1}J'_{2}$ and
$N_{2} = (\cosh\eta)^{-1}J'_{1}$ respectively.  Then, in the
infinite-$\eta$ limit,

\begin{equation}\label{eq4}
N_{1} = K_{1} - J_{2} ,\qquad N_{2} = K_{2} + J_{1} .
\end{equation}
These operators satisfy the commutation relations
\begin{equation}\label{e2like}
[J_{3}, N_{1}] = iN_{2} ,  \qquad [J_{3}, N_{2}] = -iN_{1} , \qquad
[N_{1}, N_{2}] = 0 .
\end{equation}
$J_{3}, N_{1}$, and $N_{2}$ are the generators of the $E(2)$-like little
group for massless particles.

\begin{figure} 
\centerline{\epsfig{figure=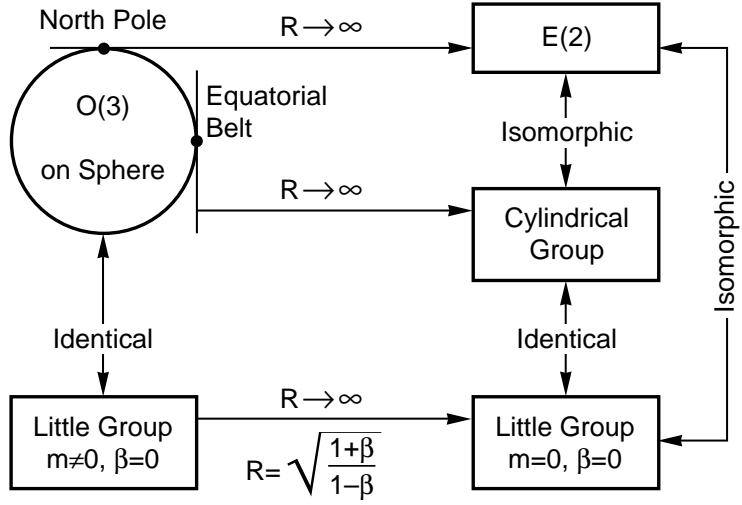,angle=0,width=100mm}}
\caption{Contraction of O(3) to E(2) and to the cylindrical group,
and contraction of the O(3)-like little group to the E(2)-like
little group.  The correspondence between E(2) and the E(2)-like
little group is isomorphic but not identical.  The cylindrical group
is identical to the E(2)-like little group.  The Lorentz boost of
the O(3)-like little group for a massive particle is the same as
the contraction of O(3) to the cylindrical group.} \label{isomor}
\end{figure}


\begin{figure} [thb]  
\centerline{\epsfig{figure=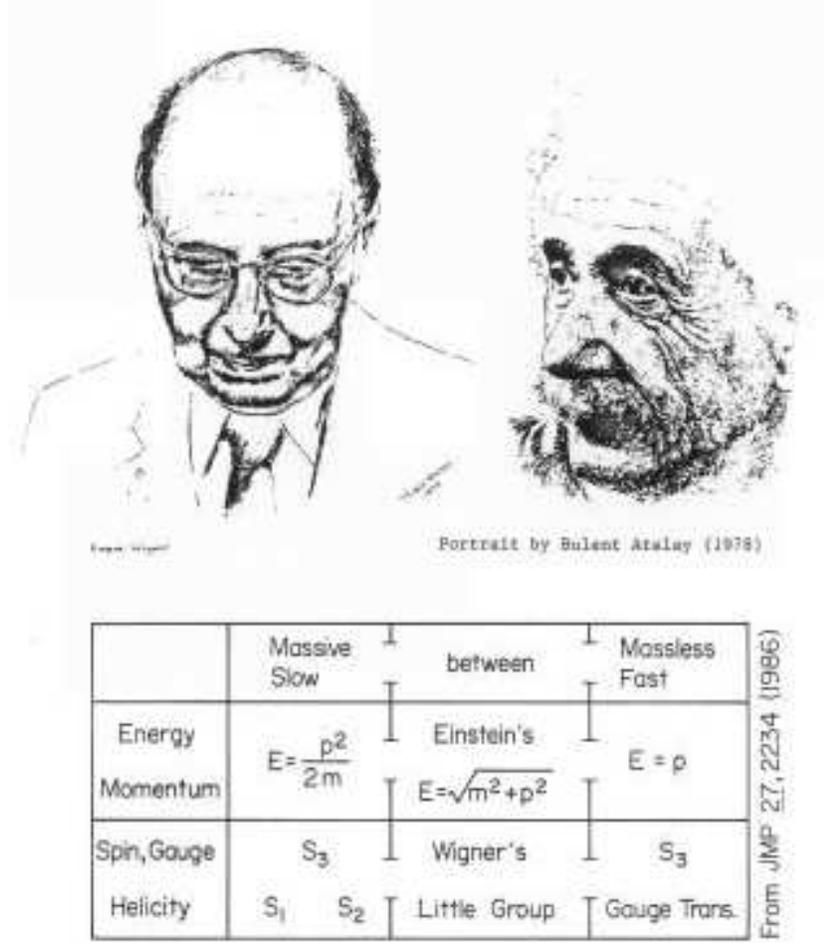,angle=0,width=110mm}}
\vspace{-5mm}
\caption{Symmetries of massive and massless particles.  Wigner's
little group unifies the internal space-time symmetries for massive and
massless particles.  We are interested in this paper how the symmetries
of spin-1/2 particles can be unified.}\label{einwig}

\end{figure}

In order to relate the above little group to transformations more
familiar to us, let us consider the two-dimensional $xy$ coordinate
system.  We can make rotations around the origin and translations along
the $x$ and $y$ axes.  The rotation generator $L_{z}$ takes the form
\begin{equation}
L_{z} = - i\left\{x {\partial \over \partial y} -
y {\partial \over \partial x} \right\}
\end{equation}
The translation generators are
\begin{equation}
P_{x} = -i {\partial \over \partial x} , \qquad
P_{y} = -i {\partial \over \partial y} .
\end{equation}
These generators satisfy the commutation relations:
\begin{equation}\label{e2com}
[L_{z}, P_{x}] = i P_{y},  \qquad [L_{z}, P_{y}] = -iP_{x} \qquad
[P_{x}, P_{y}] = 0 .
\end{equation}
These commutation relations are like those given in Eq.(\ref{e2like}).
They become identical if $L_{z}$, $P_{x}$ and $P_{y}$ are replaced by
$J_{1}$, $N_{2}$ and $N_{3}$ respectively.

This group is not discussed often in physics, but is intimately related
to our daily life.  When we drive on the streets, we make translations
and rotations, and thus make transformations of this $E(2)$ group.
Football players also make $E(2)$ transformations in their fields.

As we shall see in the following section, the translation-like degrees
of freedom in the $E(2)$-like little group lead to gauge transformations.
Then Wigner's ilttle group is like Einstein's $E = mc^{2}$
which unifies the energy-momentum relations for both massive and
massless particles.  The little group unifies the internal
space-time symmetries of massive and massless particles, as summarized
in Fig.~\ref{einwig}.  The longitudinal rotation remains as the
helicity degrees of freedom during the boost.  However, the Lorentz
boosted transverse rotations become gauge degrees of freedom, as
summarized in Fig.~\ref{einwig}.

The content of this table has been thoroughly studied in the
literature~\cite{kiwi90}.  Wigner's original paper was published in
1939~\cite{wig39}.  He worked out there the little groups for massive
and massless particles.  The task of obtaining the $E(2)$-like little
group started in the 1953 paper by Inonu and Wigner~\cite{ino53}, and
the task was not completed until 1990~\cite{kiwi90}.  What was known in
in 1953 was the Euclidean group applicable to a flat surface, but
the solution to the problem required also the cylindrical group hidden
in the four-by-four Lorentz-transformation matrices.  This has
been thoroughly studied in the literature, and Fig~\ref{isomor} gives
a complete picture of the contraction procedure.

\section{Symmetry of Massless Particles}\label{massless}
The internal space-time symmetry of massless particles is governed by
the cylindrical group which is locally isomorphic to two-dimensional
Euclidean group~\cite{kiwi90}.  In this case, we can visualize a circular
cylinder whose axis is parallel to the momentum.  On the surface of this
cylinder, we can rotate a point around the axis or translate along the
direction of the axis.  The rotational degree of freedom is associated
with the helicity, while the translation corresponds to a gauge
transformation in the case of photons.

The question then is whether this translational degree of freedom is
shared by all massless particles, including neutrinos and gravitons.
The purpose of this note is to show that the requirement of invariance
under this symmetry leads to the polarization of neutrinos~\cite{hks82}.
Since this translational degree of freedom is a gauge degree of freedom
for photons, we can extend the concept of gauge transformations to all
massless particles including neutrinos.

If we use the four-vector convention $x^{\mu} = (x, y, z, t)$, the
generators of rotations around and boosts along the $z$ axis take the
form
\begin{equation}\label{eq2}
J_{3} = \pmatrix{0&-i&0&0\cr i&0&0&0\cr 0&0&0&0\cr 0&0&0&0} , \qquad
K_{3} = \pmatrix{0&0&0&0\cr 0&0&0&0 \cr 0&0&0&i \cr 0&0&i&0} ,
\end{equation}
respectively.  The remaining four generators are readily available in
the literature~\cite{knp86}.  They are applicable also to the
four-potential of the electromagnetic field or to a massive vector
meson.

The role of $J_{3}$ is well known~\cite{wig57}.  It is the helicity
operator and generates rotations around the momentum.  The $N_{1}$ and
$N_{2}$ matrices take the form~\cite{knp86}
\begin{equation}\label{eq6}
N_{1} = \pmatrix{0&0&-i&i\cr 0&0&0&0 \cr i&0&0&0 \cr i&0&0&0} , \qquad
N_{2} = \pmatrix{0&0&0&0 \cr 0&0&-i&i \cr 0&i&0&0 \cr 0&i&0&0} .
\end{equation}
The transformation matrix is
\begin{eqnarray}
&{}& D(u,v) = \exp{\left\{-i\left(uN_{1} +
             vN_{2}\right)\right\}} \nonumber \\[1ex]
&{}& = \pmatrix{1 & 0 & -u & u \cr 0 & 1 & -v & v    \cr
 u & v & 1 - (u^{2}+ v^{2})/2 & (u^{2} + v^{2})/2 \cr
u & v & -(u^{2} + v^{2})/2 & 1  + (u^{2} + v^{2})/2} .
\end{eqnarray}
If this matrix is applied to the electromagnetic wave propagating
along the $z$ direction
\begin{equation}
A^{\mu}(z,t) = (A_{1}, A_{2}, A_{3}, A_{0}) e^{i\omega (z - t)} ,
\end{equation}
which satisfies the Lorentz condition $A_{3} = A_{0}$,
the $D(u,v)$ matrix can be reduced to~\cite{hks82}
\begin{equation}\label{eq9}
D(u,v) = \pmatrix{1&0&0&0 \cr 0&1&0&0\cr u&v&1&0\cr u&v&0&1} .
\end{equation}
While $A_{3} = A_{0}$, the four-vector $(A_{1}, A_{2}, A_{3}, A_{3})$
can be written as
\begin{equation}
(A_{1}, A_{2}, A_{3}, A_{0}) =
(A_{1}, A_{2}, 0, 0) + \lambda (0, 0, \omega, \omega) ,
\end{equation}
with $A_{3} = \lambda \omega$.  The four-vector $(0, 0, \omega, \omega)$
represents the four-momentum.  If the $D$ matrix of Eq.(\ref{eq9}) is
applied to the above four vector, the result is
\begin{equation}
(A_{1}, A_{2}, A_{3}, A_{0}) =
(A_{1}, A_{2}, 0, 0) + \lambda'(0, 0, \omega, \omega) ,
\end{equation}
with $\lambda' = \lambda + (1/\omega)\left(uA_{1} + vA_{3}\right)$.
Thus the $D$ matrix performs a gauge transformation when applied to
the electromagnetic wave propagating along the $z$
direction~\cite{hks82,janner71,hk81ajp}.

With the simplified form of the $D$ matrix in Eq.(\ref{eq9}), it is
possible to give a geometrical interpretation of the little
group~\cite{hks86}.
If we take into account of the rotation around the $z$ axis, the most
general form of the little group transformations is $R(\phi)D(u,v)$,
where $R(\phi)$ is the rotation matrix.  The transformation matrix is
\begin{equation}
R(\phi)D(u,v) = \pmatrix{\cos\phi &-\sin\phi &0&0 \cr
\sin\phi & \cos\phi &0&0 \cr u&v&1&0 \cr u&v&0&1} .
\end{equation}
Since the third and fourth rows are identical to each other,
we can consider the three-dimensional space $(x, y, z, z)$.  It is
clear that $R(\phi)$ performs a rotation around the $z$ axis.  The $D$
matrix performs translations along the $z$ axis.  Indeed, the internal
space-time symmetry of massless particles is that of the circular
cylinder~\cite{kiwi90}.

\section{Massless Spin-1/2 Particles}\label{spinhalf}
The question then is whether we can carry out the same procedure for
spin-1/2 massless particles.  We can also ask the question of whether
it is possible to combine two spin 1/2 particles to construct a
gauge-dependent four-potential.  With this point in mind, let us go back
to the commutation relations of Eq.(\ref{eq1}).  They are invariant under
the sign change of the boost operators.  Therefore, if there is a
representation of the Lorentz group generated by $J_{i}$ and $K_{i}$,
it is possible to construct a representation with $J_{i}$ and $-K_{i}$.
For spin-1/2 particles, rotations are generated by
\begin{equation}
J_{i}= {1\over 2}\sigma_{i},
\end{equation}
and the boosts by
\begin{equation}
K_{i} = (+){i\over 2}\sigma_{i}, \quad  or \quad
K_{i} = (-){i\over 2}\sigma_{i}.
\end{equation}
The Lorentz group in this representation is often called $SL(2,c)$.

If we take the (+) sign, the $N_{1}$ and $N_{2}$ generators are
\begin{equation}
N_{1} = \pmatrix{0&i \cr 0&0} ,\qquad N_{2} = \pmatrix{0&1\cr 0&0}.
\end{equation}
On the other hand, for the (-) sign, we use the ``dotted representation''
for $N_{1}$ and $N_{2}$:
\begin{equation}
\dot{N}_{1} = \pmatrix{0&0 \cr -i & 0} , \qquad
\dot{N}_{2} = \pmatrix{0&0\cr 1&0}.
\end{equation}
There are therefore two different $D$ matrices:
\begin{equation}
D(u,v) = \pmatrix{1 & u - iv \cr 0&1} , \qquad
\dot{D}(u,v) = \pmatrix{1&0 \cr -u - iv&1}.
\end{equation}
These are the gauge transformation matrices for massless spin-1/2
particles~\cite{hks82,knp86}.

As for the spinors, let us start with a massive particle at rest, and
the usual normalized Pauli spinors $\chi_{+}$ and $\chi_{-}$ for the
spin in the positive and negative $z$ directions respectively.  If we
take into account Lorentz boosts, there are two additional spinors.
We shall use the notation $\chi_{\pm}$ to which the boost generators
$K_{i} = (+){i\over 2}\sigma_{i}$ are applicable, and $\dot{\chi}_{\pm}$
to which $K_{i} = (-){i\over 2}\sigma_{i}$ are applicable.  There are
therefore four independent spinors~\cite{hks82,knp86}.  The $SL(2,c)$
spinors are gauge-invariant in the sense that
\begin{equation}\label{eq16}
D(u,v) \chi_{+} = \chi_{+} , \qquad
\dot{D}(u,v) \dot{\chi}_{-} = \dot{\chi}_{-} .
\end{equation}
On the other hand, the $SL(2,c)$ spinors are gauge-dependent in the sense
that
\begin{eqnarray}
&{}& D(u,v) \chi_{-} = \chi_{-} + (u - iv)\chi_{+} , \nonumber \\[1ex]
&{}& \dot{D}(u,v) \dot{\chi}_{+} = \dot{\chi}_{+} - (u + iv)\dot{\chi}_{-} .
\end{eqnarray}
The gauge-invariant spinors of Eq.(\ref{eq16}) appear as
neutrinos in the real world.  The Dirac equation for massless neutrinos
contains only the gauge-invariant $SL(2,c)$ spinors.

\section{The Origin of Gauge Degrees of Freedom}\label{gauge}
However, where do the above gauge-dependent spinors stand in the
physics of spin-1/2 particles?  Are they really responsible for the gauge
dependence of electromagnetic four-potentials when we construct a
four-vector by taking a bilinear combination of spinors?

The relation between the $SL(2,c)$ spinors and the four-vectors
has been discussed in the literature for massive
particles~\cite{knp86,naima64,gel63}.  However, is it
true for the massless case?  The central issue is again the gauge
transformation.  The four-potentials are gauge dependent, while the
spinors allowed in the Dirac equation are gauge-invariant.  Therefore,
it is not possible to construct four-potentials from the Dirac spinors.
However, it is possible to construct the four-vector with the four
$SL(2,c)$ spinors~\cite{hks82,knp86}.  Indeed,
\begin{eqnarray}
&{}&-\chi_{+}\dot{\chi}_{+} = (1, i, 0, 0) ,\hspace{10mm}
\chi_{-}\dot{\chi}_{-} = (1, -i, 0, 0) , \nonumber \\[1ex]
&{}& \chi_{+}\dot{\chi}_{-} = (0, 0, 1, 1) , \hspace{10mm}
\chi_{-}\dot{\chi}_{+} = (0, 0, 1, -1) .
\end{eqnarray}
These unit vectors in one Lorentz frame are not the unit vectors in other
frames. The $D$ transformation applicable to the above four-vectors is
clearly $D(u,v) \dot{D}(u,v)$.
\begin{eqnarray}
&{}& D(u,v)\dot{D}(u,v) |\chi_{+} \dot{\chi}_{+}> =
\dot{\chi} |\chi_{+}\dot{\chi}_{+}> -
(u + iv)|\chi_{+}\dot{\chi}_{-}>, \nonumber \\[1ex]
&{}& D(u,v)\dot{D}(u,v) |\chi_{-} \dot{\chi}_{-}>  =
\chi_{-}\dot{\chi}_{-}> - (u + iv) |\chi_{+}\dot{\chi}_{-}> ,\nonumber \\[1ex]
&{}& D(u,v)\dot{\chi}(u,v) |\chi _{+}\dot{\chi}_{-}> =
|\chi_{+}\dot{\chi}_{-}> .
\end{eqnarray}
The component $\chi_{-}\dot{\chi}_{+} = (0, 0, 1, -1)$ vanishes from
the Lorentz condition.
The first two equations of the above expression correspond to the gauge
transformations on the photon polarization vectors.  The third equation
describes the effect of the $D$ transformation on the four-momentum,
confirming the fact that $D(u,v)$ is an element of the little group.
The above operation is identical to that of the four-by-four $D$ matrix
of Eq.(\ref{eq9}) on photon polarization vectors.

It is possible to construct a six-component Maxwell tensor by making
combinations of two undotted and dotted spinors~\cite{knp86}.  For
massless particles, the only gauge-invariant components are
$|\chi_{+}\chi_{+}>$ and $|\dot{\chi}_{-}\dot{\chi}_{-}>$~\cite{knp86}.
They correspond to the photons in the Maxwell tensor representation with
positive and negative helicities respectively.

It is also possible to construct Maxwell-tensor fields with for a
massive particle, and obtain massless Maxwell fields by group
contraction~\cite{bask95}.  The construction of Maxwell tensors
has been thoroughly studied by Weinberg in 1964~\cite{wein64}.

\end{document}